\documentclass[conference]{IEEEtran}
\usepackage{cite}
\usepackage{graphicx}
\usepackage{epstopdf}
\graphicspath{{}}
\usepackage{graphicx}
\usepackage{amsmath}
\usepackage{stfloats}
\usepackage{amsmath,amssymb,amsfonts}
\usepackage{bm}
\usepackage{algorithmic}
\usepackage{xcolor}
\usepackage{mathtools}
\usepackage[keeplastbox]{flushend}
\usepackage{subcaption}
\begin{document}
\title{Deep Contextual Bandits for Fast Neighbor-Aided Initial Access in mmWave Cell-Free Networks}
\author{
	\IEEEauthorblockN{Insaf Ismath, Samad Ali, Nandana Rajatheva, and Matti Latva-aho}
	\IEEEauthorblockA{\textit{Center for Wireless Communication, University of Oulu, Oulu, Finland}}
	\IEEEauthorblockA{\{insaf.ismath, samad.ali, nandana.rajatheva, matti.latva-aho\}@oulu.fi}}

\maketitle

\begin{abstract}
Access points (APs) in millimeter-wave (mmWave) and sub-THz-based user-centric (UC) networks will have sleep mode functionality. 
As a result of this, it becomes challenging to solve the initial access (IA) problem when the sleeping APs are activated to start serving users. 
In this paper, a novel deep contextual bandit (DCB) learning method is proposed to provide instant IA using information from the neighboring active APs.
In the proposed approach, beam selection information from the neighboring active APs is used as an input to neural networks
that act as a function approximator for the bandit algorithm.
Simulations are carried out with realistic channel models generated using the Wireless Insight ray-tracing tool. 
The results show that the system can respond to dynamic throughput demands with negligible latency compared to the standard baseline 5G IA scheme. 
The proposed fast beam selection scheme can enable the network to use energy-saving sleep modes without compromising
the quality of service due to inefficient IA.

\end{abstract}
\IEEEpeerreviewmaketitle

\section{Introduction}

To facilitate applications like extended reality (XR) that demand high-throughput low-latency communication, networks of the future are expected to provide at least a 1000-fold increase in network throughput \cite{Qualcomm2013}.
Despite high sensitivity to blockages, millimeter wave (mmWave) and sub-THz bands will be exploited to attain throughput goals\cite{Jain2019}.
Massive multiple-input multiple-output (MIMO) enabled user-centric (UC) topology is proposed as an alternative to the cellular architecture\cite{Buzzi2017}. 
Users in a UC network are simultaneously served with coherent transmissions made by multiple APs.
Since mmWave and THz communications rely mostly on line-of-sight (LOS) links, APs are needed to be densely deployed to ensure reliability.
However, all the APs might not be required all the time to provide a good quality of service\cite{VanChien2020}.
Therefore, maintaining a network with a large number of otherwise redundant APs causes considerably inefficient power usage.

The work in \cite{Nguyen2017, VanChien2020, Feng2017} present strategies to optimize the energy consumption of the network;
each user is only served by a sub-set of APs.
The network power consumption is minimized by optimizing the AP sub-set selection to meet a minimum user throughput requirement.
The functionality of the redundant APs is reduced and put into a low power consumption mode called the \textit{sleep-state}. 
Although having the sleep-state improves energy efficiency, it compromises the agility of the network.
With highly dynamic mmWave channels, users may suddenly lose connections to some APs\cite{Jain2019}.
Furthermore, users may instantly need additional radio resources to begin or maintain high-throughput low-latency applications. 
To maintain a consistent quality of service against the above-discussed scenarios and more, sleep-state APs are required to start contributing to the network capacity as soon as the normal functionality is restored.
However, the 5G IA system achieves synchronization and identifies beams to serve users using a lengthy exhaustive search-based beam-sweep which is performed in a dedicated time-slot in the radio frame \cite{3GPP_TR38}.
Since radio frames have to be synchronized between the APs to enable simultaneous user serving, a newly restored AP has to wait for the next common IA period to perform IA.
Hence, the 5G IA method adds two latency components: lengthy IA procedure and the wait for the next IA slot, leading to a compromise between energy efficiency and quality of service.
Therefore, with the 5G initial access (IA) procedure newly restored APs are unable to instantly contribute to the network.

Recently, a new wave of interest has surged in using machine learning (ML) in wireless communications \cite{ali2020white}.
Some ML methods can be applied to the IA problem. 
For example, the work in \cite{Insaf2020} presents an ML-based approach for fast IA in ultra-dense UC systems. 
Inspired by ML solutions, in this work we propose a novel deep contextual bandit (DCB) based neighbor-aided IA approach to avoid the compromise between energy efficiency and quality of service.
With this novel approach, a newly woken AP can instantaneously start serving the users irrespective of where the AP was restored in the radio frame.
The DCB model in each AP learns about the environment through the beam choices made by the neighboring APs. 
Using this knowledge, the DCB model learns to map a given set of beam choices made by neighbors to one of its beams.
Strictly speaking, the DCB model takes the beam choice made by its neighboring APs to serve a user as the input and predicts the beam which it should select to serve the same user as the output. 

The proposed method provides instantaneously beam selection for recently restored APs irrespective of the location at the radio frame.
Hence, the energy efficiency is enhanced by enabling the use of a sleeping mechanism for APs without incurring a latency penalty, and therefore, the network flexibility is not compromised.
The performance of this neighbor-aided IA system is evaluated using realistic scenarios generated using a ray-tracing tool.
Simulation results for the considered scenario show that the instantaneous predictions of the DCB model achieve 96\% of the best case signal-to-noise ratio (SNR) while maintaining a negligible latency.

The rest of this paper is organized as follows. 
Section \ref{sec:sys_model} explains the system model used in this work and introduces the IA problem.
Section \ref{sec:p_approach} formally presents the proposed approach and provides a quick primer on DCB.
Section \ref{sec:sim} explains details about the simulations and presents numerical results and Section \ref{sec:conclution} concludes the paper.

\underline{Notations}: $(.)^T$, and $(.)^H$ denote transpose and Hermitian transpose, respectively.
$\mathcal{R}(\mathbf x)$ and $\mathcal{I}(\mathbf x)$ represents the real and imaginary parts of $\mathbf x$, respectively.
$||\mathbf x||$ and $|\mathbf x|$ denote the euclidean norm and cardinality of $ \mathbf x$, respectively.

\begin{figure}[t]
	\centering
	\includegraphics[width=0.8\linewidth]{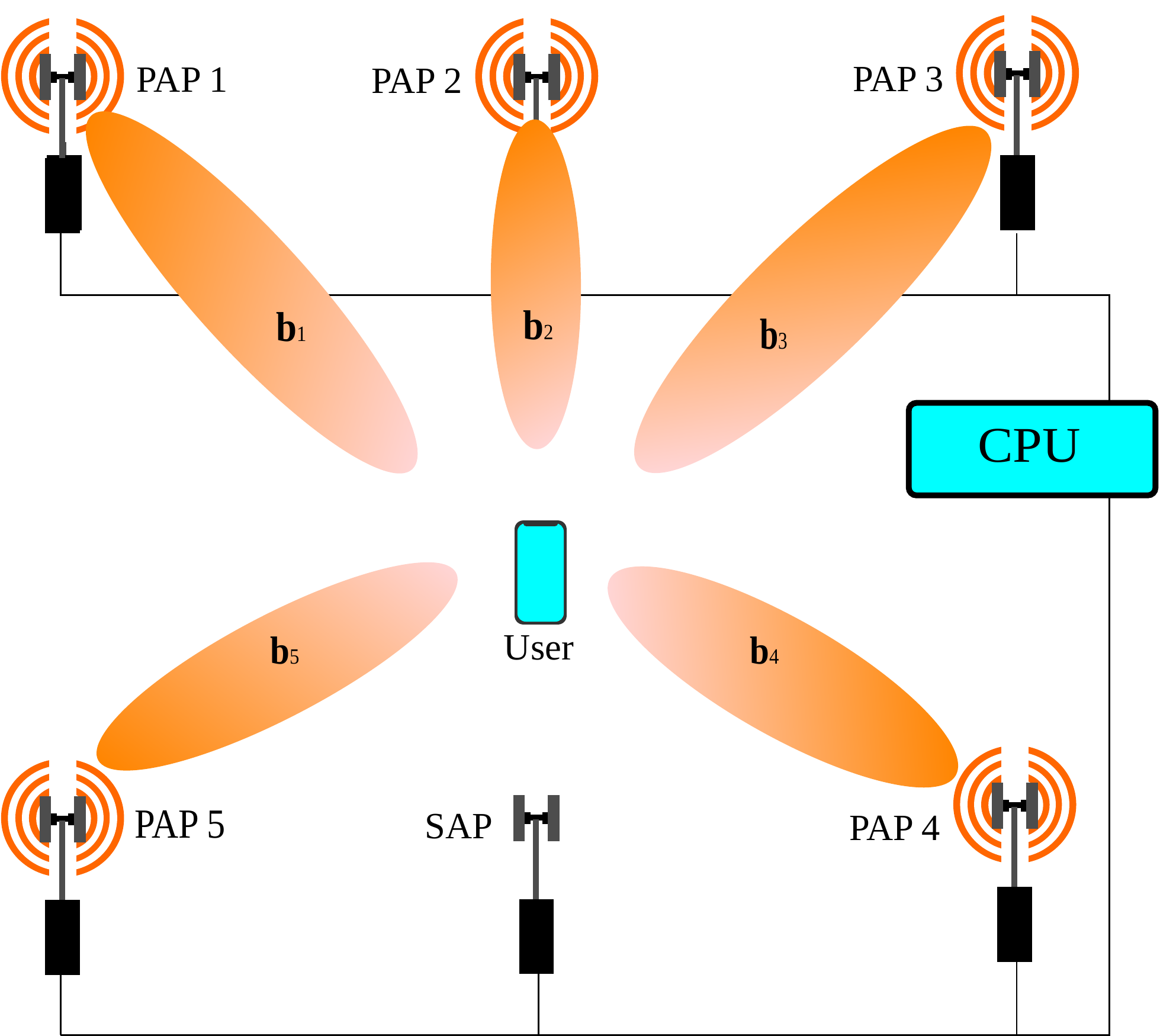}
	\caption{The system model of a UC network where a subset of APs would simultaneously serve a single user\cite{Buzzi2017}. 
	APs can either operate as awake-state PAPs or as sleep-state SAPs.	}	
	\label{fig:systemmodel}
\end{figure}

\section{System Model and Problem Formulation}
\label{sec:sys_model}

\subsection{System Model}

Consider a UC network with a set of $N$ densely deployed APs and one user as presented in Fig. \ref{fig:systemmodel}.
APs are equipped with uniform rectangular planar antenna arrays (URPA) with $M$ antenna-elements.
According to the UC architecture\cite{Buzzi2017}, the single antenna user is simultaneously served by multiple APs to ensure reliability and to maintain high throughput communication.
All APs are connected to their controlling entity, the central processing unit (CPU) via backhaul links.
To conserve energy APs can operate in two power consumption modes: the fully functional \textit{awake-state} and the reduced functionality power-saving \textit{sleep-state}\cite{Feng2017}. 
APs in the awake-state and sleep-state are defined as primary APs (PAPs) and secondary APs (SAPs), respectively.
Only a subset of APs are maintained as PAPs; remaining APs are put to the sleep-state.
Each PAP periodically performs the 5G IA procedure using beam-sweeps to identify the user serving beam and to synchronize the network. 

The mmWave channel is modeled with a clustered model\cite{Akdeniz2014}.
Each channel is generated using $K$ clusters which are made combining $Q$ paths characterized by path loss, fading, and array gain at the AP.
The subscript notations $k=1,\dots,K$ and $q=1,\dots,Q$ represent cluster and path index, respectively. 
The channel from an AP to the user $\mathbf{h} \in \mathbf{C}^{1\times M}$ is presented as
\begin{equation}
\mathbf{h} = \frac{1}{\sqrt{Q}}\sum_{k=1}^{K} \sum_{q=1}^{Q} p_{k,q}r_{k,q}\mathbf{a}(\theta_{k,q},\phi_{k,q}),\label{eq:gkmn}
\end{equation}
where 
$p_{k,q}\in \mathbf{C}$ and $r_{k,q} \in \mathbf{C}$ represent the path loss and small-scale fading gain, respectively, and 
$\mathbf{a}(\theta_{k,q},\phi_{k,q}) \in \mathbf{C}^{1\times M}$ is the AP's array gain. 
Here $\theta_{k,q}$ and $\phi_{k,q}$ are the azimuth and elevation angles of arrival at the AP, respectively.
The channels used in this work are generated using a ray-tracing tool called Wireless Insight\cite{REMCOM2017} to capture realistic mmWave behaviors.

Due to the cost and power consumption reasons, the authors in \cite{Jain2019} suggest using analog beamforming techniques for mmWave MIMO enabled APs.
Hence, this work considers analog beamforming which is implemented using an array of $M$ quantized phase shifters.
Beamforming codebook $\mathbf{F}$ is the finite set of $M$ beams generated using this arrangement and the $i$th beam $\mathbf{f_i} \in \mathbf{F}$ is $\frac{1}{\sqrt{M}} \left[ e^{j\Theta_{i,1}} \; \dots \;e^{j\Theta_{i,M}}\right] ^T$.
Here $\Theta_{i,m}$ is the phase shift added to the signal emitted from the $m$th antenna-element corresponding to the $i$th beam.

\subsection{Problem Formulation}

During IA, every AP has to identify the beam that maximizes the received SNR at the user.
The SNR component provided by the $n$th PAP to the user $\Gamma_n$ is expressed as
\begin{align}
\Gamma_n =  \frac{P_{n}||\mathbf{h_n} \mathbf{b_n}||^2}{\sigma^2},
\end{align}
where
$P_{n}$ is the transmit power of $n$th PAP,
$\mathbf{h_n}$ is the channel $\mathbf{h}$ from $n$th PAP to the user,
$\mathbf{b_n}$ is the beam chosen by $n$th PAP to serve the user,  
$\sigma^2$ is the noise power at the user, and 
$\bar{N}$ is the number of PAPs serving the user.
The IA problem for the network is presented as 
\begin{equation}
\begin{aligned}
\mathop{\operatorname {maximize} } _{\mathbf{B}} \quad & \sum_{n =1}^{\bar{N}} \Gamma_n&\\
\operatorname {subject \; to} \quad & \mathbf{b_n} \in \mathbf{F} & \forall n =1,\dots,\bar{N},
\end{aligned}
\label{eq:problem}
\end{equation}
where
$\mathbf{B}=\left[ \mathbf{b_1},...,\mathbf{b_{\bar{N}}}\right]$.
The 5G IA procedure solves \eqref{eq:problem} using a beam-sweep at each PAP. 
Beam-sweeping is performed using sequential transmissions of all or part of the beams in $\mathbf{F}$ using orthogonal resources. 
The user reports the best beam from each beam sweep to the corresponding AP. 
Furthermore, 5G IA is performed in a dedicated time slot called physical broadcast channel (PBCH)\footnote{Commonly just one IA opportunity is allocated per radio frame\cite{3GPP_TR38}.}. 
Since APs in the UC architecture perform cooperative simultaneous serving, PBCHs have to be synchronized between the APs.
When an SAP has been restored the problem formulation in \eqref{eq:problem} changes and the new problem is defined as
\begin{equation}
\begin{aligned}
\mathop{\operatorname {maximize} } _{\mathbf{B}, \mathbf{b_{0}}} \quad & \sum_{n =1}^{\bar{N}} \Gamma_n + \frac{P_{0}||\mathbf{h_0} \mathbf{b_0}||^2}{\sigma^2} &\\
\operatorname {subject \; to} \quad & \mathbf{b_n} \in \mathbf{F} & \forall n =0,\dots,\bar{N},
\end{aligned}
\label{eq:problem_new}
\end{equation}
where the familiar quantities with $0$th index correspond to the quantities of the newly restored AP.
With the 5G IA method, the newly restored AP has to wait till the next PBCH to perform IA. 
Hence, although having a sleep-state improves energy efficiency, the use of conventional beam-sweep-based IA systems for this scenario introduces an additional latency component.

However, beam choices of PAPs, i.e., $\mathbf{B}$, were found when \eqref{eq:problem} was resolved using 5G IA during the previous PBCH.
This solution for $\mathbf{B}$ is valid for \eqref{eq:problem_new} since the next PBCH is yet to arrive.
Hence, finding just $\mathbf{b_0}$ is adequate to solve \eqref{eq:problem_new}.
The proposed approach solves \eqref{eq:problem_new} by predicting $\mathbf{b_0}$ for the newly restored AP with a DCB model which uses $\mathbf{B}$ and PAP location information $\mathbf{L}$ as inputs. 
Since the proposed method avoids beam-sweeps, waiting for the next PBCH is not required, and therefore, $\mathbf{b_0}$ is solved instantly irrespective of the position in the radio frame.
Hence, the proposed method enables any recently restored AP to instantaneously supplement the throughput provided by the PAPs to a user.
With the IA-related latency component eradicated, the CPU is free to adjust the network capacity to satisfy user demands in an energy-efficient manner without compromising the quality of service.

\section{DCB based Neighbor-Aided Initial Access Approach}
\label{sec:p_approach}

\subsection{Proposed Approach}

The PAPs periodically share beam choices among the neighbors.
The CPU may utilize an algorithm similar to the work presented in \cite{VanChien2020, Nguyen2017, Feng2017} to determine which SAP to be restored.
In the proposed approach, 
the restored AP approximates the achievable SNR for each $\mathbf{f_i}$ using the function $\mathbf{g}(\mathbf{L},\mathbf{B})$ which is defined as
\begin{equation}
\begin{aligned}
\mathbf{g}(\mathbf{L},\mathbf{B}): \left\lbrace \mathbf{L} \right\rbrace \times \left\lbrace \mathbf{B} \right\rbrace  \rightarrow \left\lbrace \frac{P_{0}||\mathbf{h_0} \mathbf{f_i}||^2}{\sigma^2} \right\rbrace, \forall \mathbf{f_i} \in \mathbf{F},
\end{aligned}
\label{eq:g}
\end{equation}
where $\left\lbrace \mathbf{L} \right\rbrace$ and $\left\lbrace \mathbf{B} \right\rbrace$ represents the vector space containing all possibilities for $\mathbf{L}$ and $\mathbf{B}$, respectively.
The optimum beam is selected as $\mathbf{b_0}=\mathbf{f_{i^*}}$ where 
\begin{equation}
\mathbf{i^*}=\mathop{\operatorname {argmax}} _{\mathbf{i}\in\left\lbrace 1,\dots |\mathbf{F}| \right\rbrace }  \mathbf{g}(\mathbf{L},\mathbf{B}).
\label{eq:b_0}
\end{equation}
Nevertheless, practical deployment settings experience a plethora of phenomena such as the absence of LOS paths and shadowing among others which makes the mathematical characterization of $\mathbf{g}(\mathbf{L},\mathbf{B})$ complex. 

Authors of \cite{Alrabeiah2020} use a DNN to approximate a similarly complex function which maps beams between sub-6 GHz and mmWave bands. 
Inspired by this solution, $\mathbf{g}(\mathbf{L},\mathbf{B})$ in this work is approximated using a DNN based DCB black-box function approximator.
Each AP is equipped with a DCB model which explores the environment and learns $\mathbf{g}(\mathbf{L},\mathbf{B})$ for the respective AP.

\subsection{An Introduction to Deep Contextual Bandits (DCB)}
\label{sec:dcb}
The contextual bandit is a special instance of the reinforcement learning model called multi-arm bandits (MABs)\cite{Ali2019}.
Reinforcement learning architectures have a software entity called the agent in an environment.
The agent can interact with the environment by performing actions selected from the action-space that contains all possible actions.
Each action entails a numerical reward that judges the quality of the action. 
The goal of the agent is to maximize the reward.
Over time, the agent discovers the best action which in the case of MAB is a constant.  
However, problems in domains like wireless communication are dynamic in nature and may have different best actions depending on the circumstance. 

In contextual bandits, the agent learns the best actions for a given context $\mathcal{X}$ which characterize the state of the environment; 
in the MAB case, only one context exists and hence, the best action is a constant.
The agent can select an action by either referring to a simple context-action-reward table or using advanced algorithms like Thompson sampling\cite{THOMPSON1933}.  
After receiving the reward for the performed action, the parameters of the selection mechanism are updated. 
However, in problems where the context space is continuous and large, neither maintaining tables nor running these algorithms is feasible, and therefore, a deep neural network (DNN) is used to predict the reward for actions given $\mathcal{X}$ \cite{Mnih2015}.
Contextual bandit models with DNNs are called DCB.
The agent follows a decaying $\epsilon$-greedy exploration policy where the state-action-reward relationships are learned by exploring the environment by taking random actions with a probability of $\epsilon$.
Otherwise, the agent leverages the already acquired knowledge to choose actions to maximize reward.
Initially, $\epsilon$ is set to 1 and in every training episode, it decays by a factor of $\epsilon_{dec}$ until $\epsilon=\epsilon_{min}$.

Since the DNN has to cater to two types of inputs: location data from $\mathbf{L}$ and beam choices from $\mathbf{B}$, the context is ingested through two independent input layers.  
Hence, the context is defined as 
\begin{equation}
\mathcal{X} \coloneqq \left[
\begin{matrix}
\mathbf{L}\\
\left[\mathcal{R}\left( \mathbf{B}\right) ,  \mathcal{I}\left( \mathbf{B}\right) \right]
\end{matrix}	\right].
\end{equation}
Most neural network tools only support real numbers, hence real and imaginary parts of $\mathbf{B}$ are split and concatenated. 
The two rows of $\mathcal{X}$ corresponds to the two input layers of the DNN.
The outputs from these two input layers are concatenated and fed into the rest of the DNN.
The action-space of the agent corresponds to the beam codebook, hence each action represents a beam.
Ideally, the reward should only represent the throughput provided to the user by the AP using the beam choice made by the agent.
However, since multiple APs coherently serve the user, measuring individual contributions is not straightforward.
The reward $\alpha$ is defined as
\begin{equation}
\alpha \coloneqq \frac{ ||\mathbf{h_0}\mathbf{b_0}||^2}{\sigma^2} \times \dfrac{1}{||\mathbf{h_0}\mathbf{h_0}^H||^2}\label{eq:norm_reward}.
\end{equation}
The normalization with $||\mathbf{h_0}\mathbf{h_0}^H||^2$ term in eq. \eqref{eq:norm_reward} benchmarks the performance of the chosen beam against the theoretical optimum beam $\mathbf{h_0}^H$\cite{Goldsmith2005}.
Instantaneous $\mathbf{h_0}$ is approximated using the last available channel estimate $\mathbf{\tilde{h}_0}$ which is assumed to be in the same channel coherence interval with $\mathbf{h_0}$, and therefore, $\mathbf{h_0} \approx \mathbf{\tilde{h}_0}$. 

The performance of the DCB agent is evaluated using regret $\delta$ that benchmarks the chosen action against the action which provides the highest reward $\alpha_{max}$.
In this work, $\delta$ is defined as
\begin{equation}
\delta \coloneqq \alpha_{max}-\alpha.
\end{equation}

\section{Simulations}
\label{sec:sim}
\subsection{Simulation Environment}
\begin{figure}[t]
	\centering
	\includegraphics[trim={0 9cm 0 8.5cm},clip,width=1\linewidth]{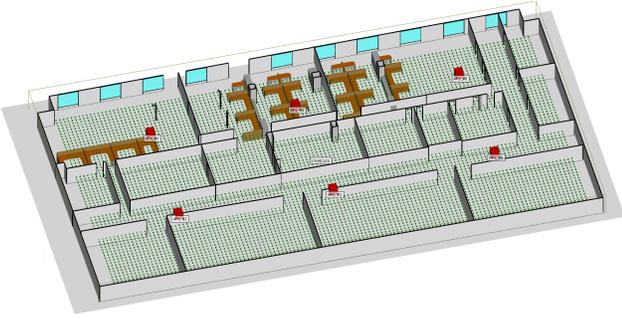}
	\caption{The indoor office simulation environment.}	
	\label{fig:sim_env}
\end{figure}

The indoor office shown in Fig. \ref{fig:sim_env} is considered for the simulations.
An area of 60 m $\times$ 30 m is partitioned using dry-wall into smaller office spaces. 
The ceiling and the floor are built with concrete.
This environment is modeled using ITU 60 GHz compliant material models provided with Wireless Insite ray-tracer\cite{REMCOM2017}.
A set of 6 APs is located 10 m apart in a grid formation centering the ceiling which is 2.6 m above the floor.
A grid of possible user locations is defined spanning the office area.
All AP and possible user locations are denoted in red and green color markings in Fig. \ref{fig:sim_env}.

A carrier frequency of 60 GHz and a channel bandwidth of 100 MHz is considered.
All APs are equipped with URPA antennas with 16 elements in the 4 $\times$ 4 configuration. 
The gain and the noise figure of every antenna are set to 5 and 3 dB, receptively.
The transmit power of the APs, and the users are set to 20 and 5 dBm, respectively.
Each AP in the awake-state and sleep-state is assumed to consumes 1 W\cite{Feng2017} and 0.01 W, respectively.
The beam codebook is comprised of 16 beams.

A simulation round consists of 50,000 episodes. 
At the beginning of each simulation round, a fixed number of PAPs are chosen at random. 
The optimization of the PAP selection criterion is not considered since it is out of the scope of this work. 
At each episode, the user is placed randomly in the grid of possible user locations.

The DNN of the DCB model has 3 hidden layers with each containing 100 neurons. 
Input layer one intakes $\mathbf{L}$ where each entry is comprised of latitude and longitude value, and therefore, it has $2\times\bar{N}$ neurons.
Input layer two ingests  $\mathbf{B}$ and it has $2\times M\times\bar{N}$ neurons.
The output layer contains $M$ neurons corresponding to $M$ beams (actions). 
ReLu and Adam have been used as the layer activation function and the optimizer respectively. 
Both dropout and learning rates are set to 0.01.

The performance of the proposed approach is analyzed in terms of regret, network capacity, and energy efficiency.

\subsection{Simulation Results}

\begin{figure}[t]
	\centering
	\includegraphics[width=1\linewidth]{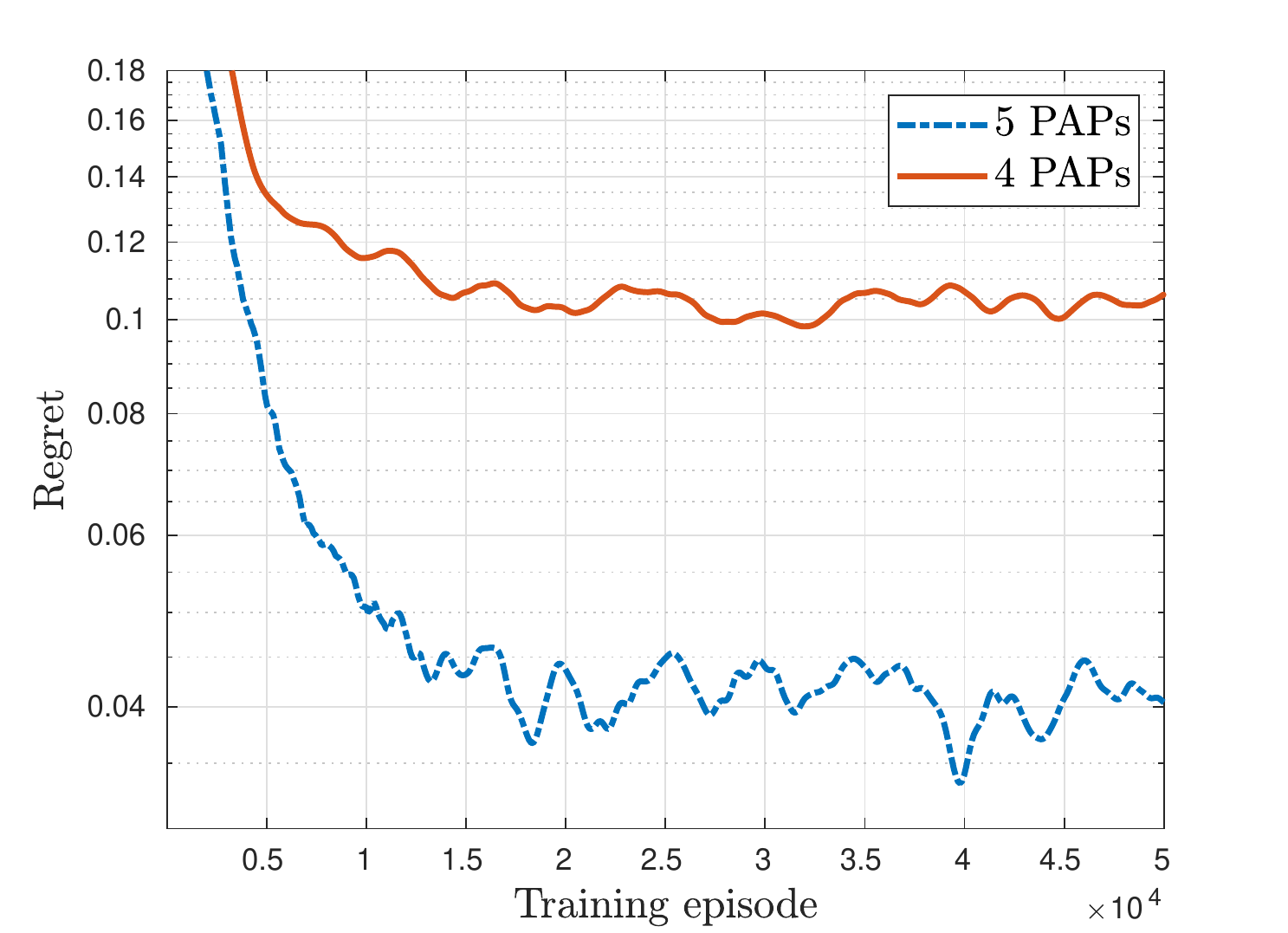}
	\caption{Regret incurred by the contextual bandit model during training.}	
	\label{fig:regret}
\end{figure}

To analyze regret, the DCB in a single SAP is trained in two scenarios where the number of PAPs is set to 4 and 5, respectively. 
Fig. \ref{fig:regret} presents the regret incurred by the DCB agent over the training episodes.
Due to higher number of neighbor in the 5 PAP case, the DCB model has access to more information compared to the 4 PAP case. 
Therefore, the DCB model in the 5 PAP case could learn faster and better compared to the 4 PAP case and results in a regret measure around 4\%.

\begin{figure*}
	\centering
	\begin{subfigure}[b]{0.49\linewidth}
		\centering
		\includegraphics[width=0.94\linewidth]{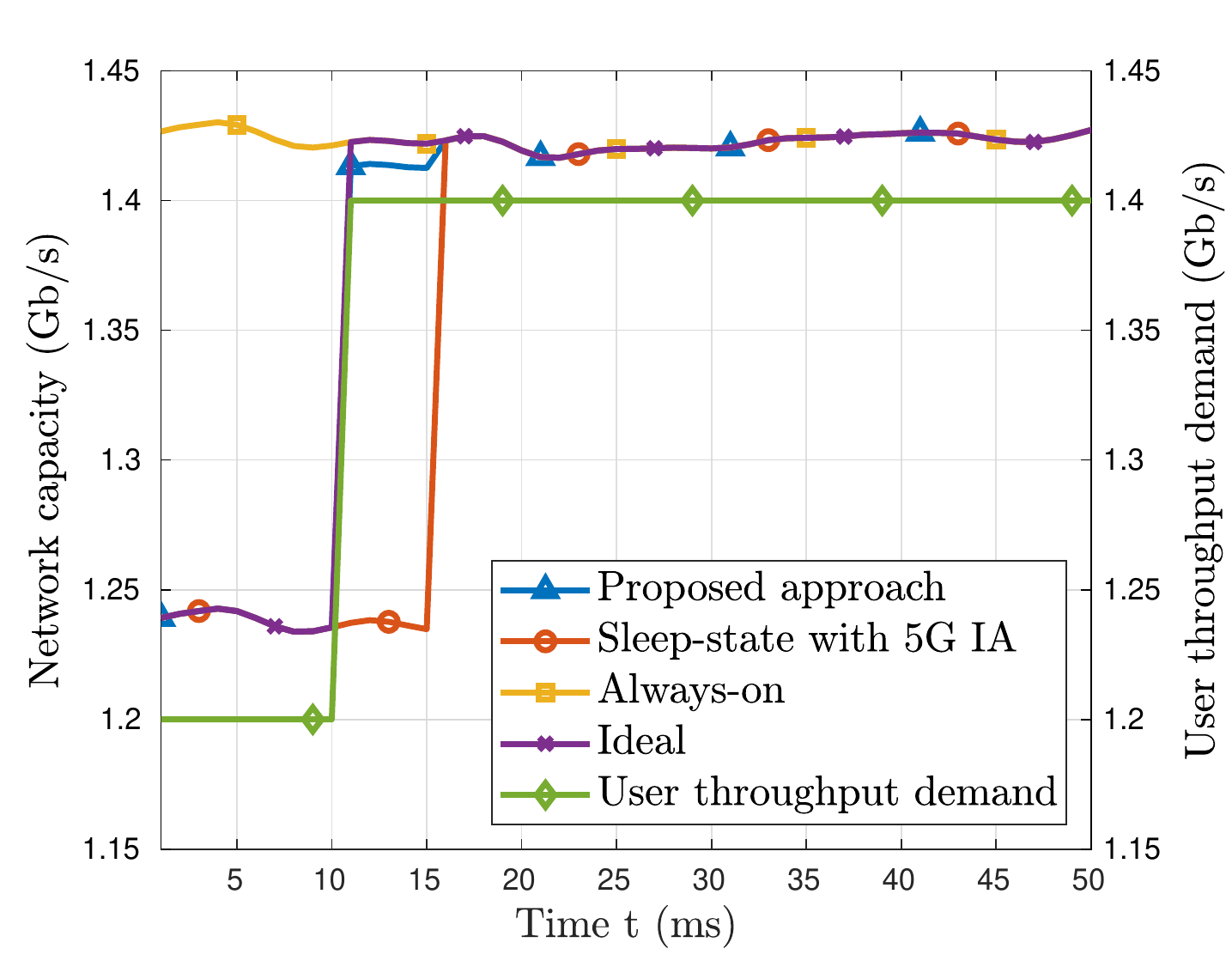}
		\caption{Network capacity.}	
		\label{fig:scenario}
	\end{subfigure}~
	\hfill
	\begin{subfigure}[b]{0.49\linewidth}
		\centering
		\includegraphics[width=1\linewidth]{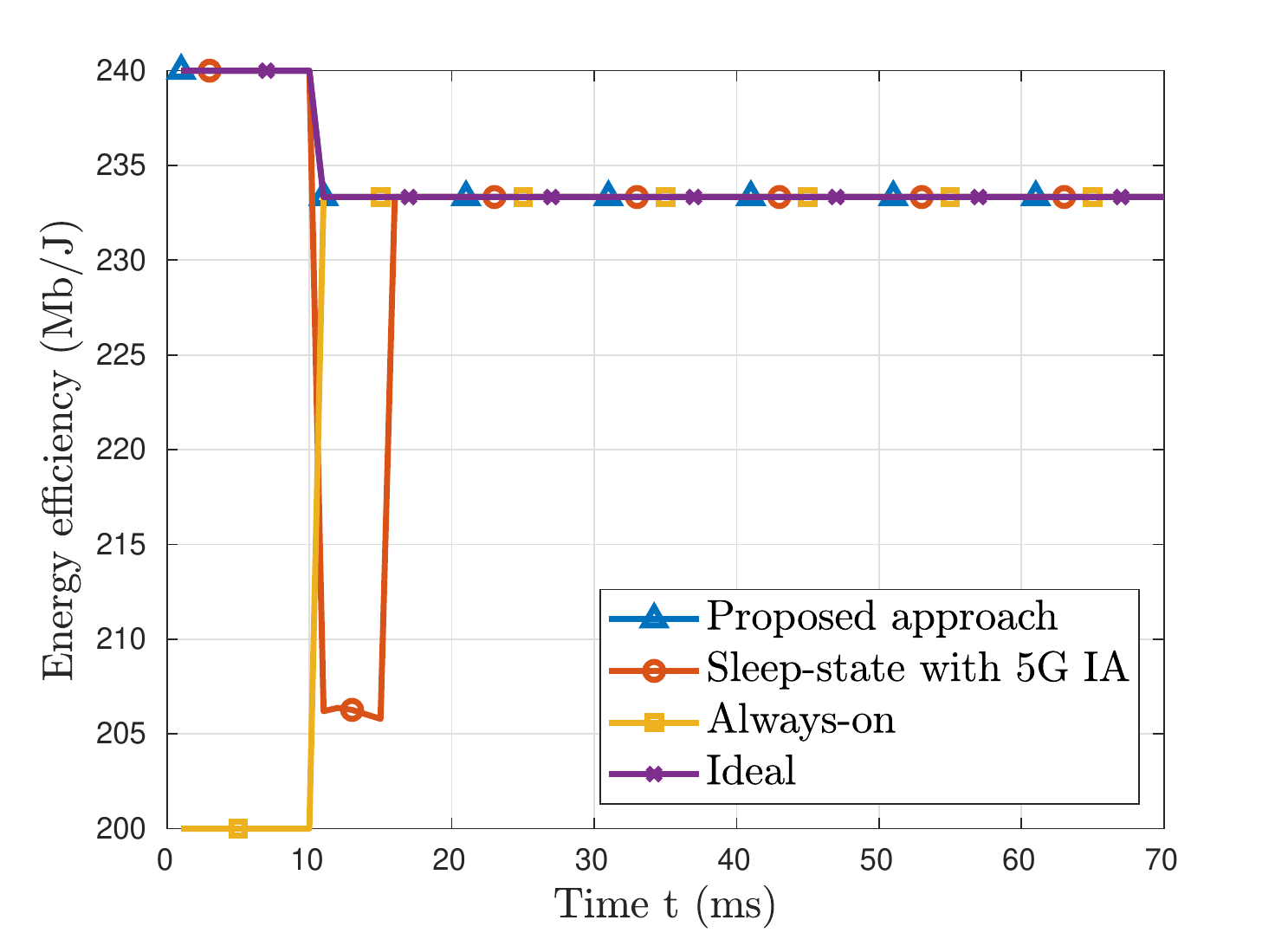}
		\caption{Energy efficiency.}	
		\label{fig:scenario_ee}
	\end{subfigure}
	\caption{Performance of proposed method, conventional sleep-state enabled algorithm, always-on approach, and ideal genie-aided system under dynamic user throughput demand. Markers in these figures are used merely to differentiate between the plots, and hence, do not represent specific simulation points.}
	\label{fig:scenario_all}

\end{figure*}

In Fig. \ref{fig:scenario_all}, a scenario capturing system performance under changing user throughput demand is investigated. 
Fig. \ref{fig:scenario} and \ref{fig:scenario_ee} present network capacity and energy efficiency, respectively.
Here the user requests for additional throughput at $t = 10$ ms. 
The user is served by 5 and 6 APs to meet the throughput demand before and after the $t = 10$ ms mark, respectively.
The performance of the proposed approach is compared with several IA schemes: 
sleep-state enabled conventional 5G IA, always-on approach, and ideal genie-aided system.

The ideal system knows the future demand and maintains the network capacity accordingly.
The always-on approach keeps all APs only in the awake-state.
Although always-on is robust against sudden changes in the user throughput demand, it has the least overall energy efficiency. 
The user may need throughput at any point in the 10 ms long radio frame\cite{T.S.G.R.A.N.3GPP2020b}.
Assuming only one beam-sweeping slot is available per radio frame, the user experiences a latency of 5 ms on average to get additional throughput from the newly restored AP using 5G IA.
With a trained DCB model, the proposed approach enables the newly restored AP to performs IA instantaneously to start providing throughput with negligible processing delay.
The proposed approach and the 5G IA both have similar power consumption profiles since they both turn on the SAP at the same time. 
Nevertheless, overall energy efficiency is higher in the proposed system compared to 5G IA since the proposed system enables the AP to contribute to the network instantaneously.
Although the analyzed case only has 5 PAPs and an SAP, the proposed method can be easily extended to other configurations.

\section{Conclusion}
\label{sec:conclution}

In this paper, we have proposed a neighbor-aided approach based on DCB to provide instantaneous IA to APs waking up from a sleep mode. First, we have defined the function which maps beam choices made by the neighbors of an AP to its beam codebook. Then, we have used a deep learning-based contextual bandit approach to solve the IA mapping problem. Finally, we have carried out simulations using realistic channels generated using the Wireless Insight raytracing tool. The results show that the DCB model can successfully learn the mapping function and provide instantaneous IA. This work, therefore, presents a faster way to provide IA using neighbor information for APs in dense mmWave-based UC networks.


\bibliographystyle{IEEEtran}
\bibliography{PhD-NeighbourAided.bib} 

\begin{thebibliography}{10}
\providecommand{\url}[1]{#1}
\csname url@samestyle\endcsname
\providecommand{\newblock}{\relax}
\providecommand{\bibinfo}[2]{#2}
\providecommand{\BIBentrySTDinterwordspacing}{\spaceskip=0pt\relax}
\providecommand{\BIBentryALTinterwordstretchfactor}{4}
\providecommand{\BIBentryALTinterwordspacing}{\spaceskip=\fontdimen2\font plus
\BIBentryALTinterwordstretchfactor\fontdimen3\font minus
  \fontdimen4\font\relax}
\providecommand{\BIBforeignlanguage}[2]{{%
\expandafter\ifx\csname l@#1\endcsname\relax
\typeout{** WARNING: IEEEtran.bst: No hyphenation pattern has been}%
\typeout{** loaded for the language `#1'. Using the pattern for}%
\typeout{** the default language instead.}%
\else
\language=\csname l@#1\endcsname
\fi
#2}}
\providecommand{\BIBdecl}{\relax}
\BIBdecl

\bibitem{Qualcomm2013}
\BIBentryALTinterwordspacing
Qualcomm, ``{The 1000x Mobile Data Challenge},'' \emph{White paper}, no.
  November, pp. 1--38, 2013. [Online]. Available:
  \url{https://www.qualcomm.com/media/documents/files/1000x-mobile-data-challenge.pdf}
\BIBentrySTDinterwordspacing

\bibitem{Jain2019}
I.~K. Jain, R.~Kumar, and S.~S. Panwar, ``{The Impact of Mobile Blockers on
  Millimeter Wave Cellular Systems},'' \emph{IEEE Journal on Selected Areas in
  Communications}, vol.~37, no.~4, pp. 854--868, 2019.

\bibitem{Buzzi2017}
S.~Buzzi and C.~D'Andrea, ``{Cell-free massive MIMO: User-centric approach},''
  \emph{IEEE Wireless Communications Letters}, vol.~6, no.~6, pp. 706--709,
  2017.

\bibitem{VanChien2020}
T.~{Van Chien}, E.~Bjornson, and E.~G. Larsson, ``{Joint power allocation and
  load balancing optimization for energy-efficient cell-free massive mimo
  networks},'' \emph{IEEE Transactions on Wireless Communications}, vol.~19,
  no.~10, pp. 6798--6812, 2020.

\bibitem{Nguyen2017}
L.~D. Nguyen, T.~Q. Duong, H.~Q. Ngo, and K.~Tourki, ``{Energy Efficiency in
  Cell-Free Massive MIMO with Zero-Forcing Precoding Design},'' \emph{IEEE
  Communications Letters}, vol.~21, no.~8, pp. 1871--1874, 2017.

\bibitem{Feng2017}
M.~Feng, S.~Mao, and T.~Jiang, ``{BOOST: Base station on-off switching strategy
  for green massive MIMO HetNets},'' \emph{IEEE Transactions on Wireless
  Communications}, vol.~16, no.~11, pp. 7319--7332, 2017.

\bibitem{3GPP_TR38}
{T. S. G. RAN}, ``{3GPP TR 38.912 (2020-07): Study on New Radio (NR) access
  technology-Rel.16},'' 3GPP, Tech. Rep., 2020.

\bibitem{ali2020white}
\BIBentryALTinterwordspacing
S.~Ali, W.~Saad, and N.~Rajatheva, ``{6G White Paper on Machine Learning in
  Wireless Communication Networks},'' \emph{University of Oulu}, 2020.
  [Online]. Available: \url{http://arxiv.org/abs/2004.13875}
\BIBentrySTDinterwordspacing

\bibitem{Insaf2020}
\BIBentryALTinterwordspacing
I.~Ismath, K.~B.~S. Manosha, S.~Ali, N.~Rajatheva, and M.~Latva-aho, ``{Deep
  Contextual Bandits for Fast Initial Access in mmWave Based User-Centric
  Ultra-Dense Networks},'' in \emph{IEEE Vehicular Technology Conference},
  2021. [Online]. Available: \url{https://arxiv.org/abs/2009.06974}
\BIBentrySTDinterwordspacing

\bibitem{Akdeniz2014}
M.~R. Akdeniz, Y.~Liu, M.~K. Samimi, S.~Sun, S.~Rangan, T.~S. Rappaport, and
  E.~Erkip, ``{Millimeter wave channel modeling and cellular capacity
  evaluation},'' \emph{IEEE Journal on Selected Areas in Communications},
  vol.~32, no.~6, pp. 1164--1179, 2014.

\bibitem{REMCOM2017}
REMCOM, ``{Wireless insite reference manual},'' 2008.

\bibitem{Alrabeiah2020}
M.~Alrabeiah and A.~Alkhateeb, ``{Deep Learning for mmWave Beam and Blockage
  Prediction Using Sub-6 GHz Channels},'' \emph{IEEE Transactions on
  Communications}, vol.~68, no.~9, pp. 5504--5518, 2020.

\bibitem{Ali2019}
S.~Ali, H.~Asgharimoghaddam, N.~Rajatheva, W.~Saad, and J.~Haapola,
  ``{Contextual bandit learning for machine type communications in the null
  space of multi-antenna systems},'' \emph{IEEE Transactions on
  Communications}, vol.~68, no.~2, pp. 1284--1296, 2020.

\bibitem{THOMPSON1933}
\BIBentryALTinterwordspacing
W.~R. THOMPSON, ``{On the likelihood that one unknown probability exceeds
  another in view of the evidence of two samples},'' \emph{Biometrika},
  vol.~25, no. 3-4, pp. 285--294, dec 1933. [Online]. Available:
  \url{https://academic.oup.com/biomet/article-lookup/doi/10.1093/biomet/25.3-4.285}
\BIBentrySTDinterwordspacing

\bibitem{Mnih2015}
V.~Mnih \emph{et~al.}, ``{Human-level control through deep reinforcement
  learning},'' \emph{Nature}, vol. 518, no. 7540, pp. 529--533, feb 2015.

\bibitem{Goldsmith2005}
A.~Goldsmith, \emph{Wireless Communications}.\hskip 1em plus 0.5em minus
  0.4em\relax Cambridge University Press, 2005.

\bibitem{T.S.G.R.A.N.3GPP2020b}
{T. S. G. RAN}, ``{3GPP TS 38.213 (2020-12): NR Physical layer procedures for
  control},'' 3GPP, Tech. Rep., 2020.

\end{thebibliography}

\end{document}